\documentclass[aps,prl,reprint,amsmath,amssymb,superscriptaddress,noeprint,nofootinbib]{revtex4-2}

\usepackage{graphicx}
\usepackage{dcolumn}
\usepackage{bm}
\usepackage{amsmath}
\usepackage{amsfonts}
\usepackage{physics}
\usepackage{xcolor}
\usepackage{caption}
\usepackage{subcaption}
\usepackage{tikz}
\usepackage[normalem]{ulem}
\usetikzlibrary{arrows.meta}
\usetikzlibrary{arrows.meta,decorations.markings}
\usetikzlibrary{decorations.pathreplacing, positioning}
\usetikzlibrary{positioning,arrows.meta,calc,shadows.blur}

\newlength{\particlesize}
\newcommand{\crt}[2]{{\hat{#1}^{\dagger}_{\mathbf{#2}}}}
\newcommand{\ann}[2]{\hat{#1}_{\mathbf{#2}}}
\newcommand{\crtann}[2]{{\hat{#1}^{\dagger}_{\mathbf{#2}}}  {\hat{#1}_\mathbf{#2}}}
\newcommand{\sumover}[1]{\sum_\mathbf{#1}}

\newcommand{\artanh}{\tanh^{\!-1}}
\newcommand{\artan}{\tan^{\!-1}}

\begin{document}

\preprint{APS/123-QED}

\title{A study of the dimer-trimer crossover in a driven three-component Fermi gas.}

\author{R. Slowinski}
\affiliation{School of Mathematical Sciences, University of Southampton, Highfield, Southampton SO17 1BJ, United Kingdom
}
\author{G. Servignat}
\affiliation{Université Paris Cité, CNRS, Astroparticule et Cosmologie, F-75013 Paris, France}
\author{F. Chevy}
\affiliation{Laboratoire de physique de l'Ecole Normale sup\'erieure, ENS, Universit\'e PSL, CNRS, Sorbonne Universit\'e, Universit\'e Paris Cit\'e, F-75005 Paris, France.
}%
\affiliation{Institut Universitaire de France (IUF), 75005 Paris, France}
\author{C. Lobo}
\affiliation{School of Mathematical Sciences, University of Southampton, Highfield, Southampton SO17 1BJ, United Kingdom
}

\date{\today}

\begin{abstract}
We develop an Effective Field Theory (EFT) for a system with three distinguishable atomic species and present a variational calculation of the two and three-body binding energies in vacuum and in the presence of a single Fermi sea. Specifically, we consider the case where the interaction between first two atomic species is externally driven so as to produce a (non-universal) closed-channel dimer whose coupling can be controlled independently of all other interactions. We then model the remaining interactions as a contact interaction between the dimer and a third atomic species which forms the medium. We derive analytical expressions for the dimer and trimer binding energies in vacuum and in medium, and in the latter case we predict a crossing between the dimer and trimer branches as a function of the atom-dimer scattering length, analogous to the usual polaron and molecule problem. Furthermore, we show that the position of this crossing can be controlled by varying the atom-atom coupling that results from the external drive and we discuss the implications of these findings.
\end{abstract}


\maketitle


\section{\label{sec:level1}Introduction}
The problem of a single impurity dressed by its surrounding medium has been studied extensively, starting with Landau's work in the 1930s \cite{Landau:1933asb} which described the properties of a mobile electron as it experiences repeated electron-phonon interactions within a lattice. The proposed quasi-particles, known as polarons have been applied to a number of physical systems and their properties are the core of many applications and solid state devices  
\cite{Appel1968193,alexandrov2010advances}. In recent decades, ultracold atoms have proven to be a powerful tool to study the physics of such polarons, thanks to the high degree of control that one has over parameters such as the interspecies scattering lengths, dimensionality, mass ratios and particle density \cite{massignan2014polarons,Massignan2025PolaronsSemiconductors}. A prominent example are Fermi polarons, which occur when particles are immersed in a  gas of spin-polarised fermions and are dressed by particle-hole (p-h) excitations of the surrounding Fermi sea \cite{chevy2006upa,combescot2007nsh,prokofev2008bdm,lobo2006nsp,schirotzek2009ofp,nascimbene2009pol,Ness2020ObservationGas,cetina2016ultrafast,VanHoucke2020High-precisionSeries,Etrych2025UniversalPolarons,Cui2020FermiCoexistence}. Similarly, the Bose polaron corresponds to a particle dressed by Bogoliubov excitations of a Bose-Einstein condensate (BEC) \cite{ardila2015impurity,Levinsen2015ImpurityEffect,jorgensen2016observation,hu2016bose}.  Contrary to the Fermi polaron for which the effect of interactions can be encapsulated in a contact two-body interaction \cite{Moser2017StabilityInteractions}, three-body physics, and most notably Efimov effect \cite{efimov1970energy,Braaten2007Efimov,ferlaino2011efimov,Naidon_2017} play a crucial role in the properties of the Bose polaron.   More recently, it was shown that the Bose and Fermi polarons could be interpolated by immersing the impurity in a superfluid of attractive spin-$1/2$ fermions \cite{laurent17Connecting,yi2015polarons,Nishida_2015, Pierce19few,Alhyder2020ImpuritySea,Alhyder2024ExploringEnergy,Bigue2022MeanSuperfluid,Hu2022CrossoverSuperfluid}. In this case, the fermions of the background form Cooper pairs described by the so-called BEC-BCS crossover \cite{zwerger2012BCSBEC}. In the weakly attractive limit the pairs are loosely bound and the impurity interacts with two weakly ideal Fermi seas, while in the strongly attractive regime, the superfluid can be described by a Bose-Einstein condensate (BEC) of strongly bound bosonic dimers. 

One intriguing and still partially open question concerns the uniqueness of the impurity ground state. For instance, it has been conjectured that the ground state of a single impurity immersed in a spin-polarized Fermi gas undergoes a transition from a polaron, dressed by particle–hole excitations of the medium, to a molecular state, in which the impurity binds with a particle from the Fermi sea to form a localized dimer \cite{prokofev2008bdm,Polaron-molecule_transition,Cui2020FermiCoexistence,mora2009ground,Ness2020ObservationGas}. This transition is predicted to be first order at zero temperature, while it becomes smooth at finite temperature \cite{Ness2020ObservationGas}. In the case of the BEC–BCS polaron, a smooth crossover is predicted between the polaron and a trimeron state, which can be interpreted as a bound state between the impurity and a Cooper pair \cite{Nishida_2015,yi2015polarons,Pierce19few}. In the case where the two Fermi seas do not interact, this transition becomes discontinuous \cite{Alhyder2020ImpuritySea}.

 In this work, we explore the transition between the molecule and the trimer states of a molecule immersed in a similar three-component system. For this, we use a toy-model where we consider three distinguishable species of particles. Two species ($a$, $b$) can interact by forming a closed channel dimer $d$ and are able to form a bound state. We then model the remaining interactions as a single contact interaction between the third species ($c$) and this dimer. 
 
 The advantage of this effective three-body interaction lies in its simplicity, since it allows us to bring our two-body intuition to bear on the three-body problem and greatly simplifies calculations. It can also be realized experimentally by assuming that the  scattering lengths between (i) the three atomic species are weak and (ii) that the dimer is non-universal, meaning that its properties are not governed by the values of the different atomic scattering lengths. The atom-atom coupling can then be controlled by radio-frequency or optical photo-association \cite{bauer2009control,clark2015quantum,Cetina2016UltrafastSea,Vivanco2025ThePolaron,Journeaux2026Two-BodyResonance,hammond2022tunable}.  
 
 Our first goal is to study this system in vacuum to confirm whether such an interaction is renormalizable and investigate the effect that varying the scattering length between the particles forming the dimer has on formation of trimers. We then further investigate the case where such a dimer is treated as an impurity in a medium of the remaining species and consider the effects that Pauli blocking and p-h interactions have on both the formation of trimers and dimers. This effectively serves to extend the study of the polaron-molecule crossover to the case where the polaron is itself a composite particle formed by a closed-channel interaction between two distinguishable particles. 

\begin{figure}[h]
\resizebox{\columnwidth}{!}{%


\tikzset{
  prop/.style={
    very thick,
  },
  g2vtx/.style={circle,draw=black, fill=black, inner sep=2pt},
  gADvtx/.style={circle,draw=black, fill=white, inner sep=2pt},
}

\begin{tikzpicture}[line width=1.25pt, scale=1,font=\LARGE]

\draw[solid]  (-5, 2) -- (-3, 2);
\draw[dashed] (-5, 1) -- (-3, 1);
\draw[dotted] (-5, 0) -- (-3, 0);
\draw[double] (-5,-2) -- (-3,-2);

\draw[decorate, decoration={brace, amplitude=8pt}, line width=1pt]
  (-2.5,2.5) -- (-2.5,-0.5);

\node[anchor=west] at (-2,0.8) {$=\frac{i}{E-\frac{\hbar^2 p^2}{2m}+i\epsilon}$};

\node[anchor=west] at (-2.5,-2.2) {$=\frac{i}{E-\nu_0-\frac{\hbar^2 p^2}{2m}+i\epsilon}$};


\node[g2vtx] (A) at (3,1) {};

\draw[solid] (A) -- ++(-1,1);
\draw[dashed]  (A) -- ++(-1,-1);
\draw[double] (A) -- ++( 1.5,0);

\node[anchor=west] at (5,1) {$=\, i g_2$};
\node[gADvtx] (B) at (3.5,-2) {};

\draw[double] (B) -- ++(-1, 1);
\draw[double] (B) -- ++( 1, 1);
\draw[dotted] (B) -- ++(-1,-1);
\draw[dotted] (B) -- ++( 1,-1);

\node[anchor=west] at (5,-2) {$=\, i g_{AD}$};

\end{tikzpicture}

%
}
\caption{\label{fig:Propagators} The Feynman rules for the effective Hamiltonian presented in Eq. (\ref{eq:MainHamiltonian}) which can be used to obtain the binding energy equations using the diagrammatic $T$-matrix approach. The bare $a$, $b$, $c$, $d$-particle propagators are denoted with solid, dashed, dotted and double lines respectively. Solid black dots and white dots with black outlines correspond to the interaction vertices $ig_2$ and $ig_{AD}$ respectively. }
\end{figure}
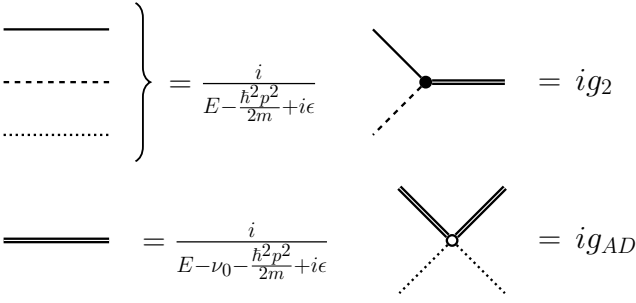

\section{General framework}
We first consider a system of three distinguishable particles which we will denote as $a$, $b$ and $c$. While $a$ and $b$ can be bosons or fermions, we will assume that $c$ is a spinless fermion. We will assume all particles are of equal mass $m$ for simplicity, though this approach can be easily generalized to unequal masses. The Hamiltonian for our effective theory ($\hat{H}$) is given by
\begin{align}\label{eq:MainHamiltonian}
    \hat{H}&=\hat{H}_0+\hat{H}_{1}+\hat{H}_{2},\\
    \hat{H}_0 &= \sumover{k}\epsilon_\textbf{k}(\crtann{a}{k}+\crtann{b}{k}+\crtann{c}{k})+\nu_\textbf{k}\crtann{d}{k},\\ 
   \hat{H}_{1}&=\frac{g_2}{\sqrt{V}}\sumover{ps}\crt{d}{p}\ann{a}{p-s}\ann{b}{s}+\crt{a}{p-s}\crt{b}{s}\ann{d}{p},\\
    \hat{H}_{2}&=\frac{g_{AD}}{V}\sumover{pst}\crt{d}{p-s}\crt{c}{s}\ann{d}{p-t}\ann{c}{t},
\end{align}
where $\crt{a}{k}$($\ann{a}{k}$) is the creation (annihilation) operator for an $a$-particle with momentum $\hbar \textbf{k}$ (likewise for $b$ and $c$), $\crt{d}{k}$ is the creation operator of a closed-channel molecule composed of one $a$ particle and one $b$ particle, $\epsilon_\textbf{k}=\hbar^2k^2/2m$ is the dispersion relation for a single particle and $\nu_\textbf{k}=\nu_0+\hbar^2k^2/4m$ is the dispersion relation for the bare molecule, with bare detuning $\nu_0$ and bare mass $2m$. Particles $a$ and $b$ can form a deeply bound closed-channel molecule $d$ thanks to the external drive which creates an atom-atom coupling $g_2$, and the newly formed molecule $d$ can then go on to interact with the third $c$-particle through a contact interaction of strength $g_{AD}$. We will take both of the interaction vertices to be constant up to some momentum cutoff $\Lambda$ which we will later take to infinity. The entire system is taken to have a volume $V$. The Feynman rules presented in Fig. \ref{fig:Propagators} can be used to compute various $T$-matrices in the two-body and three-body sectors.

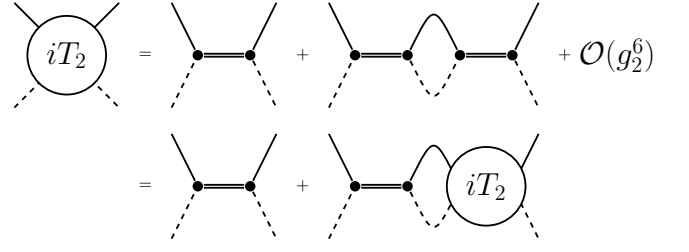
\begin{figure}[t]
\resizebox{\columnwidth}{!}{%
\begin{tikzpicture}[line width=1pt, scale=1]

\tikzset{
  blob/.style={circle, draw, minimum size=15mm, preaction={fill=white}, font=\LARGE},
  vtx/.style={circle, fill=black, inner sep=2pt}
}

\begin{scope}[shift={(0,0)}]
  \draw (-1, 1) -- ( 0,0);
  \draw[dashed] (-1, -1) -- ( 0,0);
  
  \draw (0,0) -- (1, 1);
  \draw[dashed] (0, 0) -- ( 1, -1);
  
  \node[blob] (B) at (0,0) {$iT_2$};
\end{scope}

\node at (1.5,0) {$=$};

\begin{scope}[shift={(2,0)}]
  \node[vtx] (V1) at (0.5,0) {};
  
  \draw (0, 1) -- ( V1);
  \draw[dashed] (0, -1) -- ( V1);

  \node[vtx] (V2) at (1.5,0) {};

  \draw[double] (V1)--(V2);

  \draw (V2) -- (2,1);
  \draw[dashed] (V2)--(2,-1);
\end{scope}

\node at (4.5,0) {$+$};

\begin{scope}[shift={(5,0)}]
  \node[vtx] (V1) at (0.5,0) {};
  \draw (0, 1) -- ( V1);
  \draw[dashed] (0, -1) -- ( V1);
  \node[vtx] (V2) at (1.5,0) {};

  \draw[double] (V1)--(V2);

  \node[vtx] (V3) at (2.5,0) {};
  \draw (V2) .. controls (2, 1) .. (V3);
  \draw[dashed] (V2) .. controls (2,-1) .. (V3);

  \node[vtx] (V4) at (3.5,0) {};
  \draw[double] (V3)--(V4);

  \draw (V4) -- (4,1);
  \draw[dashed] (V4)--(4,-1);  
\end{scope}

\node at (9.5,0) {$+$};

\node at (10.5,0) {\LARGE$\mathcal{O}(g_2^6)$};


\node at (1.5,-2.5) {$=$};

\begin{scope}[shift={(2,-2.5)}]
  \node[vtx] (V1) at (0.5,0) {};
  
  \draw (0, 1) -- ( V1);
  \draw[dashed] (0, -1) -- ( V1);

  \node[vtx] (V2) at (1.5,0) {};

  \draw[double] (V1)--(V2);

  \draw (V2) -- (2,1);
  \draw[dashed] (V2)--(2,-1);
\end{scope}

\node at (4.5,-2.5) {$+$};

\begin{scope}[shift={(5,-2.5)}]
  \node[vtx] (V1) at (0.5,0) {};
  \draw (0, 1) -- ( V1);
  \draw[dashed] (0, -1) -- ( V1);
  \node[vtx] (V2) at (1.5,0) {};

  \draw[double] (V1)--(V2);

  \node[vtx] (V3) at (2.5,0) {};
  \draw (V2) .. controls (2, 1) .. (V3);
  \draw[dashed] (V2) .. controls (2,-1) .. (V3);

  \node[] (V4) at (3.5,0) {};
  
  \draw (V4) -- (4,1);
  \draw[dashed] (V4)--(4,-1); 
  \node[blob] (B) at (3,0) {$iT_2$};
  
\end{scope}

\end{tikzpicture}%
}
\caption{ The Feynman diagrams representing the two-body interaction between $a$ (solid line) and $b$ (dashed line) forming a closed-channel dimer $d$ (double line) in vacuum.}
\label{fig:T2VacDiagrams}
\end{figure}

\subsection{Two-body sector in vacuum}
By taking only particles $a$ and $b$ to be present, $\hat{H}_2$ plays no role and we are left with the well-known Hamiltonian for a closed-channel interaction between $a$ and $b$ \cite{GURARIE20072, Zhai_2021}.  One can obtain the two-body binding equation by starting with the variational wave function for two particles at rest:
\begin{equation}
      \ket{\Psi_D} =\Delta_0\crt{d}{0}\ket{0}+ \sumover{k}\alpha_\textbf{k}\crt{a}{k}\crt{b}{-k}\ket{0}
\end{equation}
subject to $|\Delta_0|^2+\sumover{k}|\alpha_\textbf{k}|^2=1$. After applying the Hamiltonian to this wave function, we are able to obtain a set of coupled equations for the binding energy $E$:
\begin{align}
    (E-\nu_0)\Delta_0 &= \frac{g_2}{\sqrt{V}}\sumover{k}\alpha_\textbf{k}, \label{T2VacDelta0Eqn}\\
    (E-2\epsilon_\textbf{k})\alpha_\textbf{k} &= \frac{g_2}{\sqrt{V}}\Delta_0. \label{T2VacAlphaEqn}
\end{align}
Rearranging Eq. (\ref{T2VacDelta0Eqn}) in terms of $\Delta_0$ and substituting the result into Eq. (\ref{T2VacAlphaEqn}) yields a single equation featuring the coefficients $\alpha_\textbf{k}$. By rearranging this equation for $\alpha_\textbf{k}$ and summing over the index $\textbf{k}$, we are able to eliminate the coefficients and arrive at a self-consistent energy equation in terms of the bare parameters $g_2$ and $\nu_0$:
\begin{equation}
    \frac{E-\nu_0}{g_2^2}-\frac{1}{V}\sumover{k}\frac{1}{E-2\epsilon_k}=0. \label{T2VacBindingEqnBeforeRenormalization}
\end{equation}
In three dimensions, the result of $\sumover{k}$ is UV-divergent. This divergence can be cured by ensuring that the parameters $g_2$ and $\nu_0$ scale in a way that reproduces observables in the two-body sector. The appropriate scaling can be determined by converting the sum into an integral equation and evaluating it up to some momentum cutoff $\Lambda$, which is later taken to infinity. For this closed-channel interaction, the observables that make Eq.(\ref{T2VacBindingEqnBeforeRenormalization}) finite are $a_2$ ($ab$ scattering length) and $r_e$ ($ab$ effective range). These observables can be reproduced provided that the bare parameters satisfy
\begin{align}
    g_2^2 &= -\frac{8\pi\hbar^4}{m^2r_e},\label{effectiveRangeRenormalization}\\ 
    \frac{\nu_0}{g_2^2}&=  \left( \frac{m}{2\pi^2\hbar^2}\right) \Lambda-\left(\frac{m}{4\pi\hbar^2}\right) \frac{1}{a_2}. \label{a2Renormalization} 
\end{align}
We note that the effective range $r_e$ must be negative for Eq. \ref{effectiveRangeRenormalization} to hold. We will therefore define $R=-r_e>0$ and use this from now on for simplicity. For the same reason, we shall henceforth take $\hbar=m=1$. Eq. (\ref{T2VacBindingEqnBeforeRenormalization}) can then be recast in terms of the observable parameters as
\begin{equation}\label{T2VacEqnRenormalized}
\frac{1}{4\pi}\left[   \left( \frac{R}{2}\right)(-E)+\sqrt{-E}-\frac{1}{a_2} \right]=T_2(E)^{-1}=0,
\end{equation}
Where $T_2(E)$ is the two-body $T$-matrix in vacuum obtained by solving the integral equation depicted in Fig. \ref{fig:T2VacDiagrams}. This verifies that the values of $E$ which satisfy the self-consistent energy equation coincide with the poles of $T_2$. By Galilean invariance, we know that at finite momentum $\textbf{p}$ the poles are found by solving $T_2(E-p^2/4)^{-1}=0$. At rest we can find an analytic expression for the binding energy in terms of the observables $a_2$ and $R$ by noticing that Eq. (\ref{T2VacEqnRenormalized}) is quadratic in $\sqrt{-E}$. By defining $E=-\kappa^2$, the two possible solutions are
\begin{equation} \label{EDAnalytic}
    \kappa_\pm=\frac{1}{R}\left(-1\pm\sqrt{1+\frac{2R}{a_2}}\right).
\end{equation}
Since $R>0$, a physical bound state (for which $E$ is located on the negative real axis of the physical sheet of the complex plane) occurs if and only if $a_2>0$, in which case $\kappa_+$ corresponds to a physical bound state and $\kappa_-$ corresponds to a virtual state (a pole on the unphysical sheet). For all other values of $a_2$, neither of the solutions $\kappa_\pm$ are positive, and therefore all solutions correspond to either resonances or virtual bound states \cite{GURARIE20072, Braaten_2008}. Since we require the $ab$-dimer to be present, we restrict ourselves to the case $a_2>0$ for the remainder of this work.

\subsection{Three-body sector in vacuum}
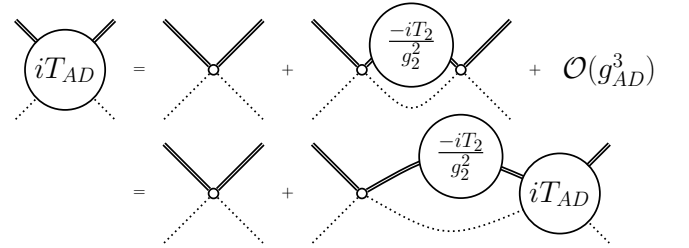
\begin{figure}[ht]
\resizebox{\columnwidth}{!}{%

\begin{tikzpicture}[line width=1pt, scale=1]

\tikzset{
  blob/.style={circle, draw, minimum size=15mm, preaction={fill=white}, font=\LARGE},
  smallBlob/.style={circle, draw, minimum size=12mm, preaction={fill=white}, font=\LARGE},
  vtx/.style={circle,draw=black, fill=white, inner sep=2pt}
}

\begin{scope}[shift={(0,0)}]
  \draw[double] (-1, 1) -- ( 0,0);
  \draw[dotted] (-1, -1) -- ( 0,0);
  
  \draw[double] (0,0) -- (1, 1);
  \draw[dotted] (0, 0) -- ( 1, -1);
  
  \node[blob] (B) at (0,0) {$iT_{AD}$};
\end{scope}

\node at (1.5,0) {$=$};

\begin{scope}[shift={(2,0)}]
  \node[vtx] (V1) at (1,0) {};
  
  \draw[double] (0, 1) -- ( V1);
  \draw[dotted] (0, -1) -- ( V1);
  
  \draw[double] (V1) -- (2,1);
  \draw[dotted] (V1)--(2,-1);
\end{scope}

\node at (4.5,0) {$+$};

\begin{scope}[shift={(5,0)}]
  \node[vtx] (V1) at (1,0) {};
  
  \draw[double] (0, 1) -- ( V1);
  \draw[dotted] (0, -1) -- ( V1);
  
  \node[vtx] (V2) at (3,0) {};

  \draw[double] (V1) .. controls (2, 1) .. (V2);
  \draw[dotted] (V1) .. controls (2,-1) .. (V2);

  \draw[double] (V2)--(4,1);
  \draw[dotted] (V2)--(4,-1);

  \node[smallBlob] (B) at (2,0.5) {$\frac{-iT_2}{g_2^2}$};

\end{scope}

\node at (9.5,0) {$+$};

\node at (11,0) {\LARGE$\mathcal{O}(g_{AD}^3)$};


\node at (1.5,-2.5) {$=$};

\begin{scope}[shift={(2,-2.5)}]
  \node[vtx] (V1) at (1,0) {};
  
  \draw[double] (0, 1) -- ( V1);
  \draw[dotted] (0, -1) -- ( V1);
  
  \draw[double] (V1) -- (2,1);
  \draw[dotted] (V1)--(2,-1);
\end{scope}

\node at (4.5,-2.5) {$+$};

\begin{scope}[shift={(5,-2.5)}]
  \node[vtx] (V1) at (1,0) {};
  
  \draw[double] (0, 1) -- ( V1);
  \draw[dotted] (0, -1) -- ( V1);

  \node[vtx] (V2) at (5,0) {};

  \draw[double] (V1) .. controls (2.75, 1) .. (V2);
  \draw[dotted] (V1) .. controls (2.75,-1) .. (V2);

  \draw[double] (V2)--(6,1);
  \draw[dotted] (V2)--(6,-1);

  \node[smallBlob] (B) at (3,0.75) {$\frac{-iT_2}{g_2^2}$};

  \node[blob] (B) at (5,0) {$iT_{AD}$};
  
\end{scope}

\end{tikzpicture}

%
}
\caption{Diagrams required to calculate $T_{AD}$ for the scattering between a dimer $d$ (double line) and $c$-particle (dotted) interacting through a contact interaction $g_{AD}$. The first expression is a perturbative expansion in $g_{AD}$ truncated after $g_{AD}^2$ and the second gives the integral equation obtained after including infinite interactions. The vacuum two-body $T$-matrix ($T_2$) enters the dimer propagator since $d$ is a composite particle.}
\label{fig:TADVac diagrams}
\end{figure}
By reintroducing the third ($c$) particle we can construct another variational wave function for three particles with zero center-of-mass (COM) momentum:
\begin{align}
    \ket{\Psi_T}&=\sumover{p}\Delta_\textbf{p}\crt{d}{p}\crt{c}{-p}\ket{0}+\sumover{kp}\Gamma_\textbf{kp}\crt{a}{\frac{p}{2}+k}\crt{b}{\frac{p}{2}-k}\crt{c}{-p}\ket{0},
\end{align}
subject to the normalization condition $\sumover{p}|\Delta_\textbf{p}|^2+\sumover{kp}|\Gamma_\textbf{kp}|^2=1$. Applying the full Hamiltonian $\hat{H}$ to $\ket{\Psi_T}$ yields the following coupled equations:
\begin{align} \label{DeltaEqn}
    \left(E-\frac{3}{2}\epsilon_\textbf{p}-\nu_0\right)\Delta_\textbf{p} &= \frac{g_2}{\sqrt{V}}\sumover{l}\Gamma_\textbf{lp}+\frac{g_{AD}}{V}\sumover{l}\Delta_\textbf{l},\\
    \left(E-\frac{3}{2}\epsilon_\textbf{p}-2\epsilon_\textbf{k}\right)\Gamma_\textbf{kp} &= \frac{g_2}{\sqrt{V}}\Delta_\textbf{p}. \label{GammaEqn}
\end{align}
Similarly to the two-body case, Eq. (\ref{GammaEqn}) can be rearranged and summed over the index $\textbf{k}$ to determine $\sumover{k}\Gamma_\textbf{kp}$, the result can then be substituted into Eq. (\ref{DeltaEqn}). Eliminating $\Delta_\textbf{p}$ then leaves us with a single equation which $E$ must satisfy
\begin{equation}\label{TADVacEqnBeforeRenormalization}
    \frac{1}{g_{AD}}=\frac{1}{V}\sumover{p}\frac{1}{E-\frac{3}{2}\epsilon_\textbf{p}-\nu_0-\frac{g_2^2}{V}\sumover{k}\frac{1}{E-\frac{3}{2}\epsilon_\textbf{p}- 2\epsilon_\textbf{k}}}.
\end{equation}
By identifying the term inside $\sumover{p}$ as $\left(T_2(E-3\epsilon_\textbf{p}/2)R/8\pi\right)$ (the renormalized dimer propagator, scaled by $R/8\pi$ and shifted by $\epsilon_\textbf{p}$ to account for the introduction of the third particle) we are able to rewrite the term in $\sumover{p}$ in terms of $\kappa_\pm$ using Eq. (\ref{EDAnalytic})  as
\begin{equation}\label{Kappa_Integrand}
   \frac{1}{\kappa_--\kappa_+}\Bigg[\frac{1}{\sqrt{\frac{3}{4}p^2-E}-\kappa_+}-\frac{1}{\sqrt{\frac{3}{4}p^2-E}-\kappa_-}\Bigg].
\end{equation}
Expressed in this way, the pole structure of the dimer propagator becomes clearer. Since $\kappa_+$ is positive and $\kappa_-$ is negative in the regime we are studying, the summation is guaranteed to be real-valued provided $E\leq-\kappa_+^2$.  Upon substituting this result into  Eq. (\ref{TADVacEqnBeforeRenormalization}) and converting it into an integral up to some finite momentum cutoff $\Lambda$ we find that the leading-order divergence ($\propto\Lambda^2$) cancels out thanks to the relative minus sign between the two terms. Nonetheless, there remains a further divergence ($\propto\Lambda^1$) which needs to be cured with the aid of a counter-term. To do this, we scale $g_{AD}$ to reproduce the atom-dimer scattering length ($a_{AD}$) when the dimer and particle $c$ scatter with zero kinetic energy, which occurs at $E=-\kappa_+^2$. The resulting scaling can be summarised as
\begin{equation} \label{gADrenormalization}
        \frac{1}{g_{AD}}=\frac{\mu}{2\pi}\frac{1}{a_{AD}}-\frac{R}{8\pi}\frac{1}{V}\sumover{p} T_2\left(-\kappa_+^2-\frac{3}{2}\epsilon_\textbf{p}\right),
\end{equation}
where $\mu=2/3$ is the reduced mass between the $c$-particle and $ab$-dimer. The resulting expression is finite and the final binding energy equation in vacuum can be written in closed form as
\begin{multline}\label{TADVacuumAnalyticExpression}
    \frac{1}{a_{AD}} = \left( \frac{2}{\pi}\right)\frac{\sqrt{2\mu}}{\kappa_+-\kappa_-}\left[ 
    (\kappa_+^2-\kappa_-^2)\log\left(\frac{\sqrt{-E}}{\kappa_+}\right) \right. \\
    -\kappa_+\sqrt{\kappa_+^2+E}\log\left(\frac{\sqrt{\kappa_+^2+E}-\kappa_+}{\sqrt{-E}}\right)\\
    +\kappa_-\sqrt{\kappa_-^2+E}\log\left(\frac{\sqrt{\kappa_-^2+E}-\kappa_-}{\sqrt{-E}}\right)\\
  \left.  -\kappa_-\sqrt{\kappa_-^2-\kappa_+^2}\log\left(\frac{\sqrt{\kappa_-^2-\kappa_+^2}-\kappa_-}{\sqrt{\kappa_+^2}}\right) \right].
\end{multline}
We remark that it is possible to bring this equation into a more familiar form in terms of the elementary function $\artan$, however such an expression is only valid for $E\leq-\kappa_-^2$. For $-\kappa_-^2<E\leq-\kappa_+^2$, one has to select the appropriate branch cut for $\sqrt{\kappa_-^2+E}$ and switch to using $\artanh$. The form presented in Eq. (\ref{TADVacuumAnalyticExpression}) is equivalent and valid for all $E\leq\kappa_+^2$. The above expression also coincides with the poles of the atom-dimer $T$-matrix $(T_{AD})$ which may be obtained by solving the integral equation presented in Fig. \ref{fig:TADVac diagrams}.

\begin{figure}[t]
\includegraphics[width=\linewidth]{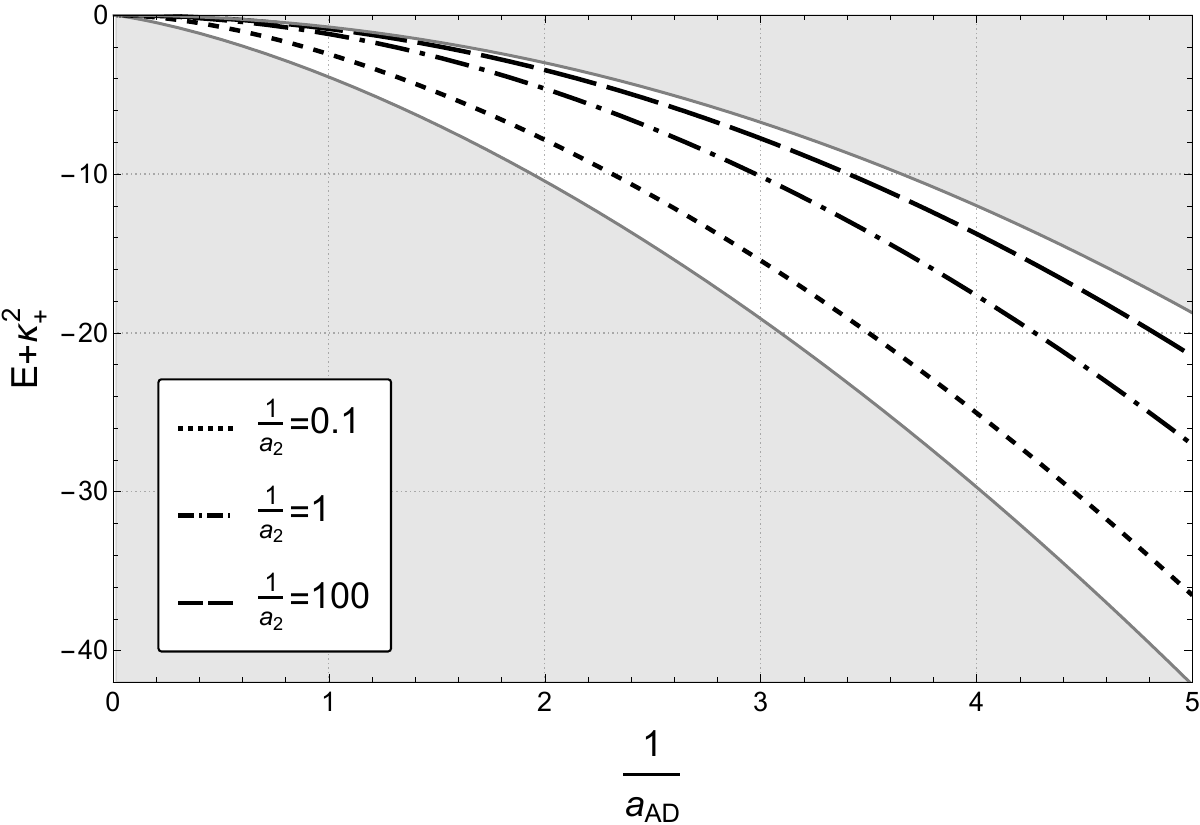}
\caption{\label{fig:TADVac Results} A graph of atom-dimer binding energy relative to the dimer binding energy ($\Delta E=E+\kappa_+^2$) in vacuum as a function of inverse atom-dimer scattering length ($1/a_{AD}$) for a selection of $ab$-scattering lengths ($a_2$). The upper bound corresponds to the point dimer limit and the lower bound corresponds to the unitary dimer limit, with the shaded gray areas indicating the regions where trimers cannot form for any combination of $a_2$ and $a_{AD}$. The remaining results have been found by numerically solving Eq. (\ref{TADVacuumAnalyticExpression}) with effective range fixed so $R=0.1$. We find that by varying $1/a_2$ the trimer solutions interpolate smoothly between these two limits.
}
\end{figure}

\subsubsection{Point dimer limit}
In the limit $R/a_2\to+\infty$ the size of the dimer tends to zero and the resulting $ab$-dimer is expected to increasingly resemble a single particle of mass $2m$. Using Eq. (\ref{EDAnalytic}) we find that in this limit $\kappa_\pm\approx\pm\sqrt{\frac{2}{Ra}}$, hence the dimer binding energy ($E_2$) is approximately
\begin{equation}
    E_2=-\kappa_+^2\approx-\frac{2}{Ra_2},
\end{equation}
and the vacuum binding energy equation approaches
\begin{equation}
    \frac{\mu}{2\pi}\frac{1}{a_{AD}}=\frac{1}{V}\sumover{p}\frac{1}{E+\kappa_+^2-\frac{1}{2\mu}\textbf{p}^2}+\frac{1}{V}\sumover{p}\frac{1}{\frac{1}{2\mu}\textbf{p}^2},
\end{equation}
which is exactly the equation one would expect to obtain when considering the scattering of two particles of unequal masses using a contact interaction, except for a shift of the rest energy by $E_2=-\kappa_+^2$. The solution of this equation is given by
\begin{equation}
    E=-\kappa_+^2-\frac{3}{4a_{AD}^2},\quad(a_{AD}>0).
\end{equation}
This result is used to obtain the upper bound for the trimer binding energy in Fig. \ref{fig:TADVac Results}. Comparing this analytic expression with the results obtained by numerically solving Eq. (\ref{TADVacuumAnalyticExpression}) for a range of values of $a_2$ we verify that for all $a_2>0$ the respective trimer binding energies must fall below this value.

\subsubsection{Unitary dimer limit}
We now turn our attention to the unitary limit where $1/a_2\to0^+$ and the binding energy of the dimer tends to zero. In this limit, we find $\kappa_+\approx 0$ and $\kappa_-\approx -2/R$. The resulting binding energy equation approaches the form
\begin{align}
    \frac{\mu}{2\pi}\frac{1}{a_{AD}}&=-\frac{1}{V}\sumover{p}\frac{1}{\frac{1}{2\mu}\textbf{p}^2-E-\kappa_-\sqrt{\frac{1}{2\mu}\textbf{p}^2-E}}\\ \notag
    &+\frac{1}{V}\sumover{p}\frac{1}{\frac{1}{2\mu}\textbf{p}^2-\kappa_-\sqrt{\frac{1}{2\mu}\textbf{p}^2}}.
\end{align}
This process eliminates the length scale $a_2$, however since our model contains the effective range term there remains we well defined length scale $R$ which ensures that this equation remains renormalizable in the three-body sector. This is reminiscent of how the effective range term helps set the ground-state trimer energy in the usual Efimov scenario with three resonantly-interacting particles, without which there is no length scale to set the three-body parameter. This equation can be written analytically as
\begin{align}
    \frac{1}{a_{AD}}&=-\frac{2\sqrt{2\mu}}{\pi}\Bigg[ -\kappa_-\log\left(\frac{\sqrt{-E}}{-2\kappa_-}\right)\\ \notag
    &+\sqrt{\kappa_-^2+E}\log \left(\frac{\sqrt{\kappa_-^2+E}-\kappa_-}{\sqrt{-E}}\right)
    \Bigg].
\end{align}
This can be solved numerically and we have used these results to determine the lower bound for the trimer energy in Fig. \ref{fig:TADVac Results}.

\section{Many-body case}

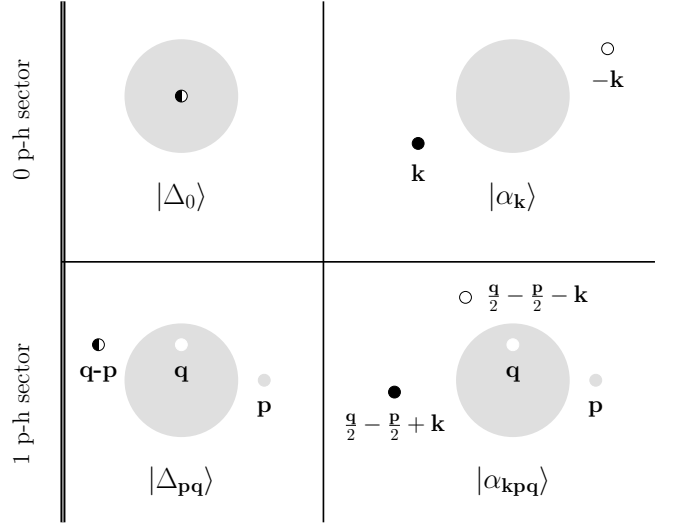
\begin{figure}[t]
\resizebox{\columnwidth}{!}{%
\begin{tikzpicture}[
  >=Latex,
  font=\Large,
  sea/.style={
    circle,
    minimum size=24mm,
    inner sep=0pt,
    fill=gray!25,
    draw=none
  },
  particleA/.style={
    circle,
    fill=black,
    draw=black,
    minimum size=\particlesize,
    inner sep=0pt
  },
  particleB/.style={
    circle,
    fill=white,
    draw=black,
    minimum size=\particlesize,
    inner sep=0pt
  },
  particleD/.style={
    circle,
    draw=black,
    fill = white,
    minimum size=\particlesize,
    inner sep=0pt,
    path picture={
      \begin{scope}
        \clip (path picture bounding box.south west)
              rectangle
              (path picture bounding box.north);
        \fill[black] (path picture bounding box.south west)
                     rectangle
                     (path picture bounding box.north east);
      \end{scope}
    }
  },
  particleC/.style={
    circle,
    fill=gray!25,
    draw=gray!25,
    minimum size=\particlesize,
    inner sep=0pt
  },
  hole/.style={
    circle,
    fill=white,
    draw=white,
    minimum size=\particlesize,
    inner sep=0pt
  }
]

\draw[double,line width=1.2pt] (-0.5,-3) -- (-0.5,8);

\draw[solid,line width=1pt] (-0.52,2.5) -- (12,2.5);
\draw[solid,line width=1pt] (5,-3) -- (5,8);

\node[rotate=90, anchor=south, font=\Large] at (-1,5.5) {0 p-h sector};
\node[rotate=90, anchor=south, font=\Large] at (-1,-0.5) {1 p-h sector};

\coordinate (TopLeft) at (2,6);
\coordinate (TopRight) at (9,6);
\coordinate (BottomLeft) at (2,0);
\coordinate (BottomRight) at (9,0);

\node[sea] (S0) at (TopLeft) {};
\node[below=0.5cm of S0, font=\LARGE] {$\ket{\Delta_0}$};

\node[particleD] at (TopLeft) {};

\node[sea] (S1) at (TopRight) {};
\node[below=0.5cm of S1, font=\LARGE] {$\ket{\alpha_\textbf{k}}$};

\node[particleA] (A1) at ([xshift=-2cm,yshift=-1cm]TopRight) {};
\node[below=2mm of A1] {$\textbf{k}$};

\node[particleB] (B1) at ([xshift=2cm,yshift=1cm]TopRight) {};
\node[below=2mm of B1] {$-\textbf{k}$};

\node[sea] (S2) at (BottomLeft) {};
\node[below=0.5cm of S2, font=\LARGE] {$\ket{\Delta_\textbf{pq}}$};

\node[particleD] (D1) at ([xshift=-1.75cm,yshift=0.75cm]BottomLeft) {};
\node[below=2mm of D1] {$\textbf{q-p}$};

\node[particleC] (C1) at ([xshift=1.75cm]BottomLeft) {};
\node[below=2mm of C1] {$\textbf{p}$};

\node[hole] (H1) at ([yshift=0.75cm]BottomLeft) {};
\node[below=2mm of H1] {$\textbf{q}$};

\node[sea] (S3) at (BottomRight) {};
\node[below=0.5cm of S3, font=\LARGE] {$\ket{\alpha_\textbf{kpq}}$};

\node[particleC] (C2) at ([xshift=1.75cm]BottomRight) {};
\node[below=2mm of C2] {$\textbf{p}$};

\node[hole] (H2) at ([yshift=0.75cm]BottomRight) {};
\node[below=2mm of H2] {$\textbf{q}$};

\coordinate (abcom) at ([xshift=-1.75cm,yshift=0.75cm]BottomRight);

\node[particleA] (A2) at ([xshift=-0.75cm,yshift=-1cm]abcom) {};
\node[below=2mm of A2] {$\frac{\textbf{q}}{2}-\frac{\textbf{p}}{2}+\textbf{k}$};

\node[particleB] (B2) at ([xshift=0.75cm,yshift=1cm]abcom) {};
\node[right=2mm of B2] {$\frac{\textbf{q}}{2}-\frac{\textbf{p}}{2}-\textbf{k}$};

\end{tikzpicture}%
}
\caption{\label{fig:Dimer Ansatz pictoral} A visual depiction of the terms included in $\ket{\Psi_D}_M$. The large gray circle at the centre of each diagram corresponds to an unperturbed $N$-particle Fermi sea of $c$-particles. Black, white (with black outline), gray and black+white dots correspond to $a$, $b$, $c$ and $d$-particles respectively. The first two terms (0 p-h sector, top) are the same in the vacuum case except for the introduction of the Fermi sea. The remaining terms (bottom) are a result of a single p-h pair interaction with the Fermi sea, with a hole of momentum $\textbf{q}$ depicted as a white dot (with no outline) inside the Fermi sea.}
\end{figure}

\begin{figure}[t]
\resizebox{\columnwidth}{!}{%
\begin{tikzpicture}[
  >=Latex,
  font=\Large,
  sea/.style={
    circle,
    minimum size=24mm,
    inner sep=0pt,
    fill=gray!25,
    draw=none
  },
  particleA/.style={
    circle,
    fill=black,
    draw=black,
    minimum size=\particlesize,
    inner sep=0pt
  },
  particleB/.style={
    circle,
    fill=white,
    draw=black,
    minimum size=\particlesize,
    inner sep=0pt
  },
  particleD/.style={
    circle,
    draw=black,
    fill = white,
    minimum size=\particlesize,
    inner sep=0pt,
    path picture={
      \begin{scope}
        \clip (path picture bounding box.south west)
              rectangle
              (path picture bounding box.north);
        \fill[black] (path picture bounding box.south west)
                     rectangle
                     (path picture bounding box.north east);
      \end{scope}
    }
  },
  particleC/.style={
    circle,
    fill=gray!25,
    draw=gray!25,
    minimum size=\particlesize,
    inner sep=0pt
  },
  hole/.style={
    circle,
    fill=white,
    draw=white,
    minimum size=\particlesize,
    inner sep=0pt
  }
]

\draw[solid,line width=1pt] (3.5,-2.5) -- (3.5,2.5);

\coordinate (Left) at (0,0);
\coordinate (Right) at (8,0);

\node[sea] (S0) at (Left) {};
\node[below=0.5cm of S0, font=\LARGE] {$\ket{\Delta_\textbf{p}}$};

\node[particleD] (D0) at ([xshift=-2cm,yshift=1cm]Left) {};
\node[below=2mm of D0] {$-\textbf{p}$};

\node[particleC] (C0) at ([xshift=2cm,yshift=-1cm]Left) {};
\node[below=2mm of C0] {$\textbf{p}$};

\node[sea] (S1) at (Right) {};
\node[below=0.5cm of S1, font=\LARGE] {$\ket{\alpha_\textbf{kp}}$};

\node[particleC] (C1) at ([xshift=2cm,yshift=-1cm]Right) {};
\node[below=2mm of C1] {$-\textbf{p}$};

\coordinate (abcom) at ([xshift=-2cm,yshift=1cm]Right);

\node[particleA] (A1) at ([xshift=-1cm,yshift=-1cm]abcom) {};
\node[below=2mm of A1] {$\frac{\textbf{p}}{2}+\textbf{k}$};

\node[particleB] (B1) at ([xshift=1cm,yshift=1cm]abcom) {};
\node[right=2mm of B1] {$\frac{\textbf{p}}{2}-\textbf{k}$};

\end{tikzpicture}%
}
\caption{\label{fig:Trimer Ansatz pictoral} A visual depiction of the terms included in $\ket{\Psi_T}_M$. The large gray circle at the centre of each diagram corresponds to an unperturbed $(N-1)$-particle Fermi sea of $c$-particles. Pauli blocking prevents particle $c$ (gray) from having momentum $\textbf{p}$ lower than the Fermi momentum $k_F$ (hence $|\textbf{p}|>k_F$). Black, white (with black outline) and black+white dots correspond to $a$, $b$ and $d$-particles respectively.}
\end{figure}
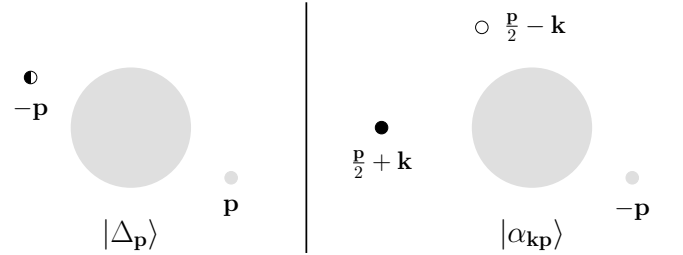

We now extend our model to include effects induced by the presence of a $c$-particle Fermi sea. We can do so by extending both of our variational wave functions
\begin{align}
      \ket{\Psi_{D}}_M &=\Delta_0\crt{d}{0}\ket{FS}_N+ 
      \sumover{k}\alpha_\textbf{k}\crt{a}{k}\crt{b}{-k}\ket{FS}_N\notag\\
      &+\sumover{pq}\Delta_\textbf{pq}\crt{d}{q-p}\crt{c}{p}\ann{c}{q}\ket{FS}_N \label{dimerAnsatzMedium}\\ 
      &+\sumover{kpq}\alpha_\textbf{kpq}\crt{a}{\frac{q}{2}-\frac{p}{2}+k}\crt{b}{\frac{q}{2}-\frac{p}{2}-k}\crt{c}{p}\ann{c}{q}\ket{FS}_N,\notag\\
      \ket{\Psi_T}_M&=\sumover{p}\Delta_\textbf{p}\crt{d}{-p}\crt{c}{p}\ket{FS}_{N-1}\notag\\ 
      &+\sumover{kp}\Gamma_\textbf{kp}\crt{a}{-\frac{p}{2}+k}\crt{b}{-\frac{p}{2}-k}\crt{c}{p}\ket{FS}_{N-1}, 
      \label{trimerAnsatzMedium}
\end{align}
where $\ket{\Psi_D}_M$ $\left(\ket{\Psi_T}_M\right)$ corresponds to the modified dimer (trimer) wave function in medium. Due to Pauli blocking, the momenta $\textbf{p}$ and $\textbf{q}$ are subject to the following boundary conditions
\begin{equation}
   |\textbf{p}|>k_F\quad,\quad |\textbf{q}|<k_F,
\end{equation}
where $k_F$ is the Fermi momentum. The state $\ket{FS}_N=\prod_{|\textbf{Q}|<k_F}\crt{c}{Q}\ket{0}$ denotes an unperturbed $N$-particle Fermi sea of $c$-particles. The two variational wave functions are constructed so that in the limit of no Fermi sea they both reduce to their vacuum counterparts, however they also have identical particle numbers for fixed $N$, which allows for direct comparison of their binding energies. Adding the Fermi sea does not change the first two terms of $\ket{\Psi_D}_M$ compared to its vacuum counterpart, however it does admit the possibility of p-h interactions, which we include up to a single p-h pair in the remaining two terms. The trimer wave function $\ket{\Psi_T}_M$ has the same form as in vacuum, however in the medium all momenta with index $\textbf{p}$ are restricted from below due to Pauli blocking by the Fermi sea. We do not extend this to include further p-h pairs since $\ket{FS}_{N-1}=c_{|\textbf{k}|=k_F}\ket{FS}_N$, implying that a hole is already present with respect to the $N$-particle Fermi sea. A visual depiction of each of the terms in these variational wave functions is presented in Fig. \ref{fig:Dimer Ansatz pictoral} for $\ket{\Psi_D}_M$ and Fig. \ref{fig:Trimer Ansatz pictoral} for $\ket{\Psi_T}_M$ respectively.

\subsection{Trimer in medium}
Applying the Hamiltonian to Eq. (\ref{trimerAnsatzMedium}) we arrive at the binding energy equation for a trimer at rest in medium. This equation is identical to Eq. (\ref{TADVacEqnBeforeRenormalization}) except we restrict the summation to $|\textbf{p}|>k_F$. Since the scaling of $g_{AD}$ is fixed by the observables in vacuum, we can use Eq. (\ref{gADrenormalization})
to rewrite the binding energy equation in terms of the observable $a_{AD}$. The resulting expression can be written analytically as
\begin{align}
    \notag \frac{1}{a_{AD}} &=\frac{2}{\pi} k_F+\frac{2\sqrt{2\mu}}{(\kappa_+-\kappa_-)\pi}\Bigg[F(\kappa_-)-F(\kappa_+) \\
    &+ (\kappa_+^2-\kappa_-^2)\log\left(\frac{\sqrt{k_F^2-2\mu E}+k_F}{\sqrt{2\mu\kappa_+^2}}\right)\\
  \notag  &-\kappa_-\sqrt{\kappa_-^2-\kappa_+^2}\log\left(\frac{\sqrt{\kappa_-^2-\kappa_+^2}-\kappa_-}{\kappa_+}\right)\Bigg],\label{TADMedAnalytic}
\end{align}
where
\begin{align}
   \notag  F(\kappa)&=\kappa\sqrt{\kappa^2+E}\Bigg[ \log \left( \sqrt{\kappa^2+E}-\kappa \right) \\
  &+  \log\left( \sqrt{2\mu(\kappa^2+E)}+k_F\right)\\
 \notag  &-  \log \left( \sqrt{\kappa^2+E}\sqrt{k_F^2-2\mu E}-\kappa k_F\right) \Bigg].\label{kappaFunction}
\end{align}
In the limit $k_F\to0$ we recover Eq. (\ref{TADVacuumAnalyticExpression}) which is the binding energy for a trimer in vacuum. Similarly, in the point-dimer limit ($1/a_2\to+\infty$) with finite $k_F$, Eq. (\ref{TADMedAnalytic}) approached the well-known binding energy equation for two particles (of unequal masses) in medium. We have solved Eq. (\ref{TADMedAnalytic}) numerically for finite $k_F$ by fixing all two-body parameters and varying $1/a_{AD}$, the resulting binding energy curves correspond to the solid black lines in Fig. \ref{fig:DimerTrimerMedium}. To support our results we consider the behaviour of Eq. (\ref{TADMedAnalytic}) in the asymptotic limits $1/a_{AD}\to\pm\infty$.

\subsubsection{Case $1/a_{AD}\to +\infty$}
In this limit the lhs of Eq. (\ref{TADMedAnalytic}) is large and positive, which can only be matched by the rhs if $E$ is large and negative. In this limit we have $E/k_F^2>>1$ and the approximation $k_F\to0$ becomes increasingly accurate. We therefore find that in this asymptotic limit the role of Pauli blocking on trimer formation becomes negligible and hence the trimer binding energy in medium approaches the trimer binding energy in vacuum (from above).

\subsubsection{Case $1/a_{AD}\to-\infty$}
In contrast with the vacuum case, we find that valid trimer solutions exist for $1/a_{AD}<0$.  Such solutions for are a purely many-body effect and correspond to a Cooper pair between a composite $ab$-dimer and a $c$-particle. Unlike for $1/a_{AD}>0$, these bound states are not localised in position space, and instead the size of the bound state can be much larger than the average interparticle spacing between the individual $c$-particles forming the medium. The reason we are able to obtain solutions for such values of $a_{AD}$ in medium but not in vacuum can be explained by examining Eq. (\ref{TADVacEqnBeforeRenormalization}). Since the term inside $\sumover{p}$ can be written as Eq. (\ref{Kappa_Integrand}) we know there is a physical pole at $E=-\kappa_+^2$, with the counterterm in Eq. (\ref{gADrenormalization}) corresponding to the result at this exact energy. These two facts lead to the rhs of Eq. (\ref{TADVacuumAnalyticExpression}) being positive for all $E\leq-\kappa_+^2$ with the the minium value of $0$ achieved exactly when $E=-\kappa_+^2$. For energies above this, one would have to integrate over a pole at $|\textbf{p}|=\sqrt{2\mu(E+\kappa_+^2)}$ which would give a complex contribution in vacuum. In medium, this pole can be present provided it does not lie on the integration region $|\textbf{p}|\geq k_F$, therefore one is able to consider higher energies provided $E<-\kappa_+^2+k_F^2/(2\mu)$. In the asymptotic limit ($1/a_{AD}\to-\infty$) the trimer binding energies tend towards this value, which can be written as
\begin{align}
    -\kappa_+^2+\frac{k_F^2}{(2\mu)}=\left(-\kappa_+^2+\frac{k_F^2}{4}\right)+\frac{k_F^2}{2},
\end{align}
where the right hand side holds due to the definition of $\mu$. Written in this form, it becomes apparent that in this limit binding energy approaches the kinetic energy of an $ab$-dimer and $c$-particle both with momenta of magnitude $k_F$, suggesting the $ab$-dimer becomes bound to a particle at the surface of the Fermi sea, forming a Cooper pair. 

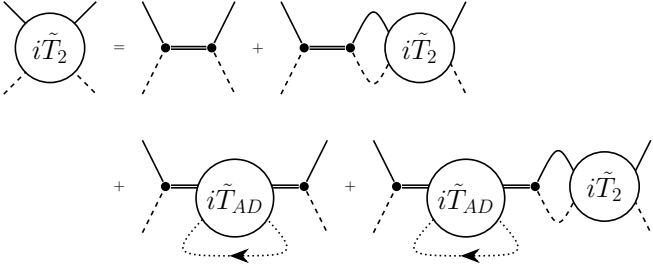
\begin{figure}[t]
\resizebox{\columnwidth}{!}{%

\begin{tikzpicture}[line width=1pt, scale=1]

\tikzset{
  blob/.style={circle, draw, minimum size=15mm, preaction={fill=white}, font=\LARGE},
  vtx/.style={circle, fill=black, inner sep=2pt}
}

\begin{scope}[shift={(0,0)}]
  \draw (-1, 1) -- ( 0,0);
  \draw[dashed] (-1, -1) -- ( 0,0);
  
  \draw (0,0) -- (1, 1);
  \draw[dashed] (0, 0) -- ( 1, -1);
  
  \node[blob] (B) at (0,0) {$i\tilde{T}_2$};
\end{scope}

\node at (1.5,0) {$=$};

\begin{scope}[shift={(2,0)}]
  \node[vtx] (V1) at (0.5,0) {};
  
  \draw (0, 1) -- ( V1);
  \draw[dashed] (0, -1) -- ( V1);

  \node[vtx] (V2) at (1.5,0) {};

  \draw[double] (V1)--(V2);

  \draw (V2) -- (2,1);
  \draw[dashed] (V2)--(2,-1);
\end{scope}

\node at (4.5,0) {$+$};

\begin{scope}[shift={(5,0)}]
  \node[vtx] (V1) at (0.5,0) {};
  
  \draw (0, 1) -- ( V1);
  \draw[dashed] (0, -1) -- ( V1);

  \node[vtx] (V2) at (1.5,0) {};

  \draw[double] (V1)--(V2);

  \node[vtx] (V3) at (2.5,0) {};
  \draw (V2) .. controls (2, 1) .. (V3);
  \draw[dashed] (V2) .. controls (2,-1) .. (V3);

  \node[] (V4) at (3.5,0) {};
  
  \draw (V4) -- (4,1);
  \draw[dashed] (V4)--(4,-1); 
  \node[blob] (B) at (3,0) {$i\tilde{T}_2$};
\end{scope}


\node at (1.5,-3) {$+$};

\begin{scope}[shift={(2,-3)}]
  \node[vtx] (V1) at (0.5,0) {};
  
  \draw (0, 1) -- ( V1);
  \draw[dashed] (0, -1) -- ( V1);

  \node[vtx] (V2) at (3.5,0) {};

  \draw[double] (V1)--(V2);

  \draw (V2) -- (4,1);
  \draw[dashed] (V2)--(4,-1);

  \draw[
  dotted,
  postaction={decorate},
  decoration={
    markings,
    mark=at position 0.52 with {\arrow{Stealth[scale=1.6]}}
  }
]
(2,0)
  .. controls (3.5,-1.5) .. (2,-1.5)
  .. controls (0.5,-1.5) .. (2,0);

  \node[blob] (B) at (2,-0.25) {$i\tilde{T}_{AD}$};

\end{scope}

\node at (6.5,-3) {$+$};

\begin{scope}[shift={(7,-3)}]
  \node[vtx] (V1) at (0.5,0) {};
  
  \draw (0, 1) -- ( V1);
  \draw[dashed] (0, -1) -- ( V1);

  \node[vtx] (V2) at (3.5,0) {};

  \draw[double] (V1)--(V2);

  \node[vtx] (V3) at (4.5,0) {};

  \draw (V2) .. controls (4, 1) .. (V3);
  \draw[dashed] (V2) .. controls (4,-1) .. (V3);

  \node[vtx] (V4) at (5.5,0) {};
  
  \draw (V4) -- (6,1);
  \draw[dashed] (V4)--(6,-1); 
  \node[blob] (B) at (5,0) {$i\tilde{T}_2$};

  \draw[
  dotted,
  postaction={decorate},
  decoration={
    markings,
    mark=at position 0.52 with {\arrow{Stealth[scale=1.6]}}
  }
]
(2,0)
  .. controls (3.5,-1.5) .. (2,-1.5)
  .. controls (0.5,-1.5) .. (2,0);

  \node[blob] (B) at (2,-0.25) {$i\tilde{T}_{AD}$};

\end{scope}

\end{tikzpicture}

%
}
\caption{\label{fig:T2Med diagrams} Diagrams used to calculate the two-body $T$-matrix between the impurities $a$ (solid) and $b$ (dashed) in medium $(\tilde{T}_2)$ up to a single p-h pair in the $c$-particle Fermi sea. The medium atom-dimer $T$-matrix ($\tilde{T}_{AD}$) enters whenever the $ab$-dimer interacts with a $c$-particle forming the medium, producing a hole which is denoted by a backwards pointing (dotted) arrow. }
\end{figure}

\begin{figure*}[htbp]
    \centering

    \begin{subfigure}{0.45\textwidth}
        \centering
        \caption{}
        \includegraphics[width=\linewidth]{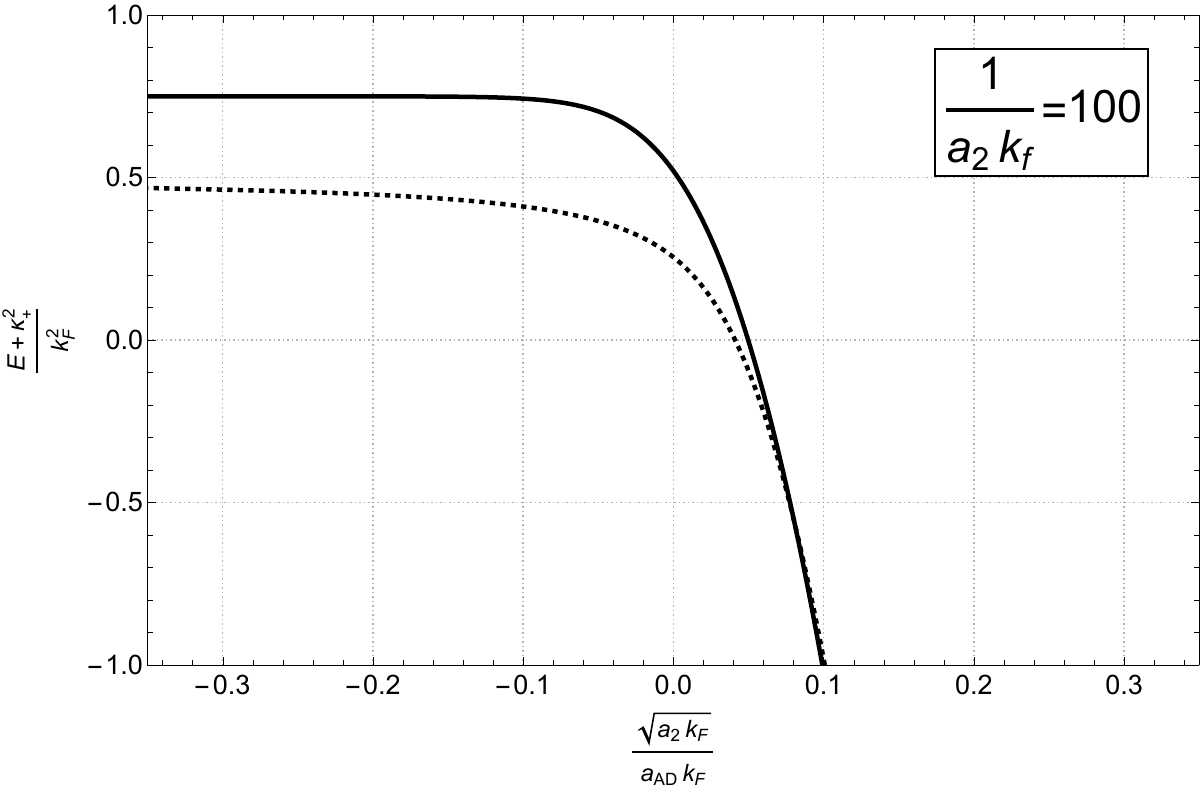}
        \label{fig:a2i100}
    \end{subfigure}
    \hfill
    \begin{subfigure}{0.45\textwidth}
        \centering
        \caption{}
        \includegraphics[width=\linewidth]{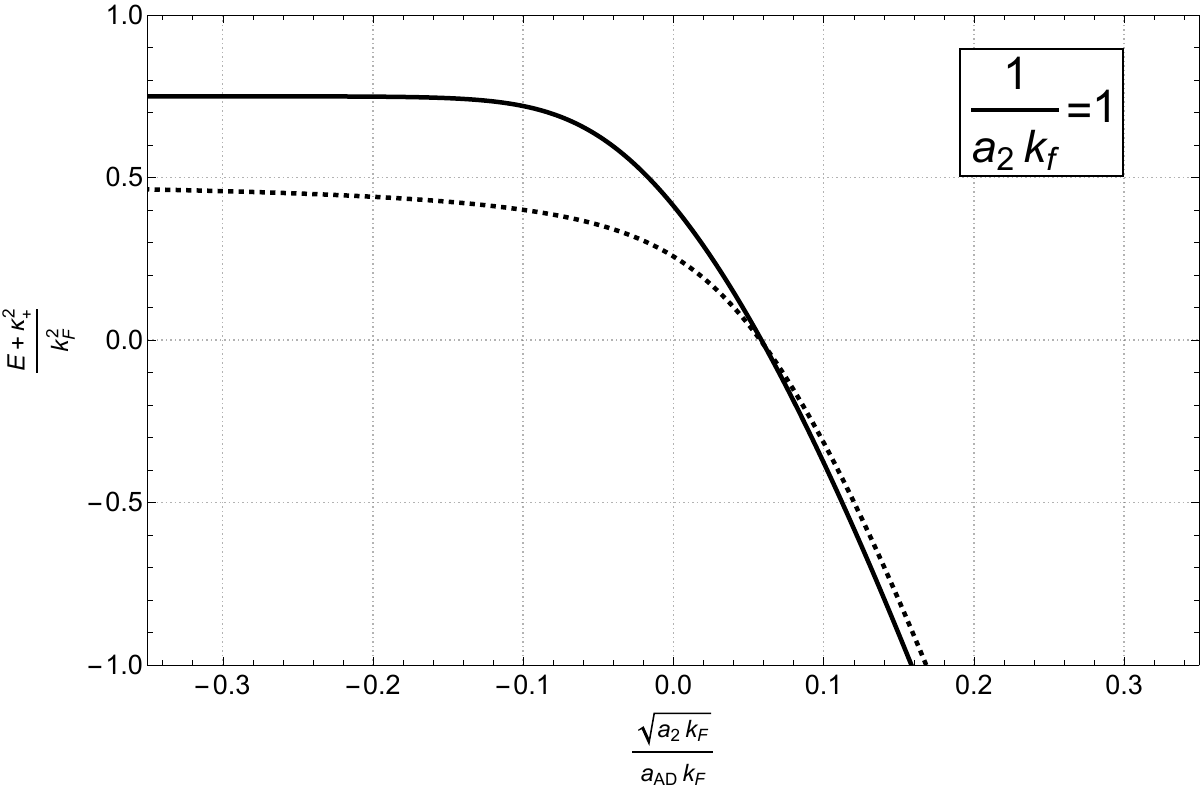}
        \label{a2i1}
    \end{subfigure}

    \medskip

    \begin{subfigure}{0.45\textwidth}
        \centering
        \caption{}
        \includegraphics[width=\linewidth]{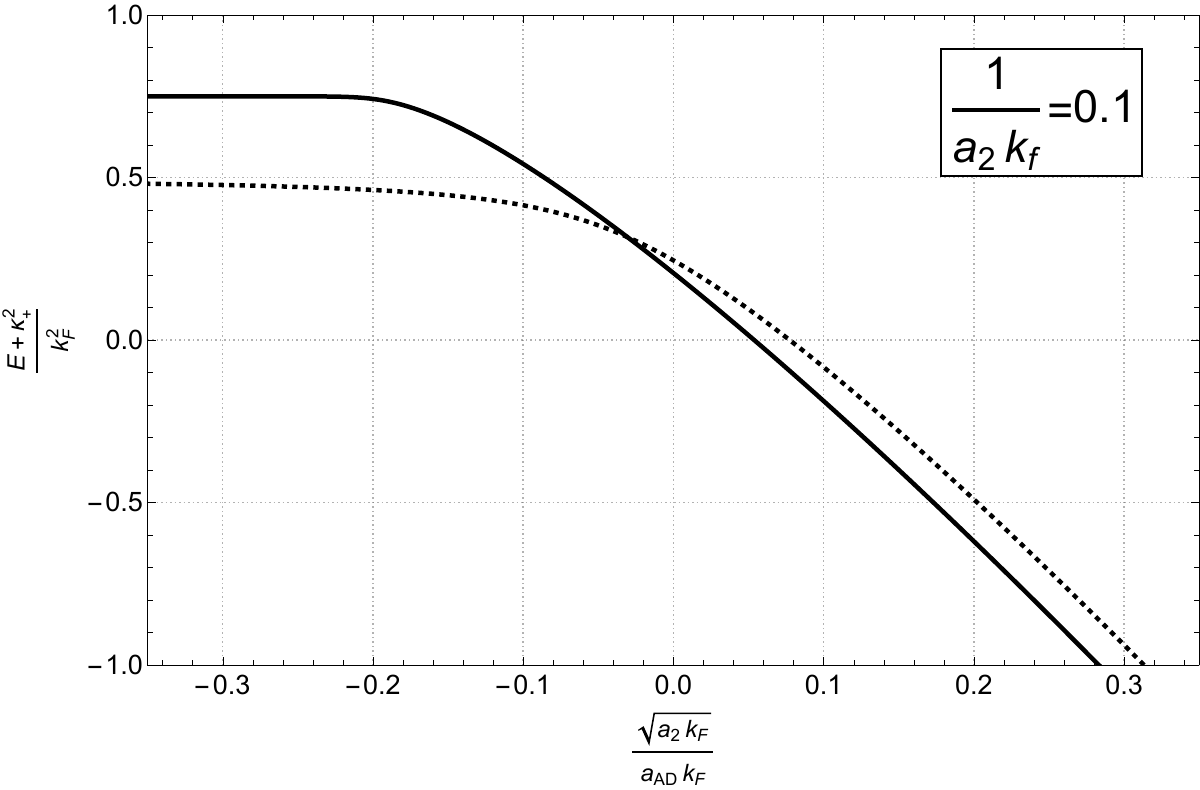}
        \label{a2i001}
    \end{subfigure}
    \hfill
    \begin{subfigure}{0.45\textwidth}
        \centering
        \caption{}
        \includegraphics[width=\linewidth]{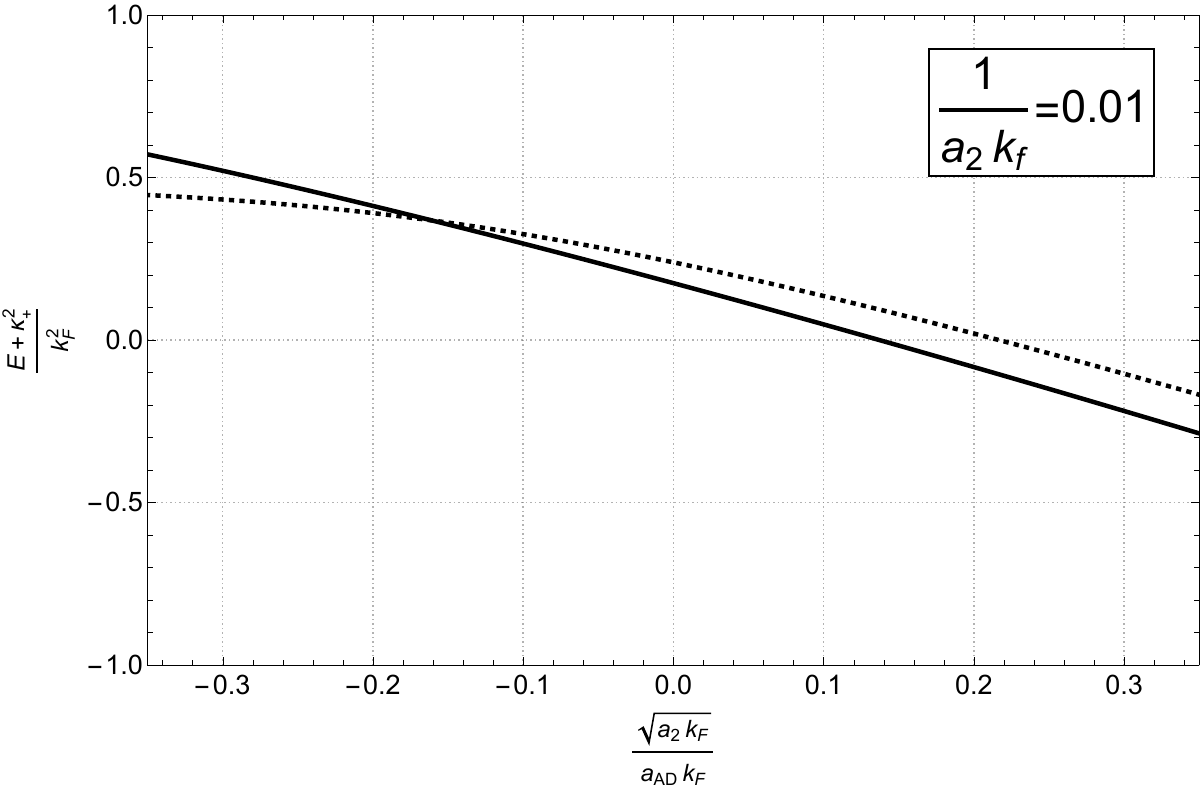}
        \label{DiffData}
    \end{subfigure}

    \caption{Comparison of the trimer binding energies (solid lines) and dimer binding energies (dotted lines) in medium as a function of $1/a_{AD}$ while keeping two-body observables fixed. For $(a_2k_F)^{-1}=(100,1,0.1,0.01)$ we observe a crossing at $\sqrt{a_2k_F/}(a_{AD}k_F)=(8.1,6.1,-3.1,-16)\times10^{-2}$. Effective range is fixed to $Rk_F=0.1$ for all figures.}
    \label{fig:DimerTrimerMedium}
\end{figure*}

\subsection{Dimer in medium}
Using the variational approach together with Eq. (\ref{dimerAnsatzMedium}) we arrive at a series of coupled equations for $E$:
\begin{align}
    (E-2\epsilon_\textbf{k})\alpha_\textbf{k} &= \frac{g_2}{\sqrt{V}}\Delta_0,\label{alphakeqn}\\ 
     (E-\nu_0)\Delta_0 &= \frac{g_2}{\sqrt{V}}\left( \sumover{k}\alpha_\textbf{k} \right)+\frac{g_{AD}}{V}\left(\sumover{q}\chi_\textbf{q}\right), \label{Delta0eqn} \\ 
   (E-\epsilon_\textbf{kpq})\alpha_\textbf{kpq}&=\frac{g_2}{\sqrt{V}}\Delta_{\textbf{pq}}, \label{alphakpqeqn} \\ 
    (E-\nu_\textbf{pq})\Delta_\textbf{pq}&=\frac{g_2}{\sqrt{V}}\left( \sumover{k}\alpha_\textbf{kpq}\right)+\frac{g_{AD}}{V}\chi_\textbf{q}, \label{Deltapqeqn} 
\end{align}
where $\epsilon_\textbf{kpq}=\epsilon_\textbf{(q-p)/2+k}+\epsilon_\textbf{(q-p)/2-k}+\epsilon_\textbf{p}-\epsilon_\textbf{q}$, $\nu_\textbf{pq}=\nu_0+\frac{1}{2}\epsilon_\textbf{q-p}+\epsilon_\textbf{p}-\epsilon_\textbf{q}$ and $\chi_\textbf{q}$ is defined as
\begin{equation}
    \label{chiDefinition} \chi_\textbf{q}=\Delta_0+\sumover{p}\Delta_{\textbf{pq}}.
\end{equation}
As we are not focusing on the comparison with the trimer binding energy at this stage, we have defined $E+\kappa_+^2=0$ to correspond to a dimer at rest in a Fermi sea of $N$ $c$-particles. When studying the competition between dimer and trimer formation we shall perform an energy shift of the dimer results by $+k_F^2/2$ to account for the additional particle in this Fermi sea compared to the trimer. These equations can be simplified by observing that the rhs of Eq. (\ref{alphakeqn}) and Eq. (\ref{alphakpqeqn}) are constant with respect to the index $\textbf{k}$. After rearranging for $\alpha_\textbf{k}$ ($\alpha_\textbf{kpq}$) and summing over $\textbf{k}$ the result can be substituted into Eq. (\ref{Delta0eqn}) (Eq. (\ref{Deltapqeqn})), reducing our problem to solving the two coupled equations
\begin{align}
    \left(E-\nu_0 -\frac{g^2_2}{V}\sumover{k}\frac{1}{E-2\epsilon_\textbf{k}}\right)\Delta_0 &= \frac{g_{AD}}{V}\left(\sumover{q}\chi_\textbf{q}\right),\label{delta0reduced}\\ \label{deltapqreduced}
    \left( E-\nu_\textbf{pq}-\frac{g_2^2}{V}\sumover{k}\frac{1}{E-\epsilon_\textbf{kpq}}\right)\Delta_\textbf{pq} &= \frac{g_{AD}}{V}\chi_\textbf{q}.
\end{align}
We can rearrange Eq. (\ref{deltapqreduced}) for $\Delta_\textbf{pq}$ and take the sum over the index $\textbf{p}$ to obtain 
\begin{equation}
    \label{sumDeltapq} \sumover{p}\Delta_{\textbf{pq}} = \chi_\textbf{q}\frac{g_{AD}}{V}\sumover{p}\frac{{1}}{ E-\nu_\textbf{pq}-\frac{g_2^2}{V}\sumover{k}\frac{1}{E-\epsilon_\textbf{kpq}}},
\end{equation}
where we have been able to factor out $\chi_\textbf{q}$ since it is independent of the index $\textbf{p}$. Usually this crucial step would not be possible in a problem involving three explicit short-range interactions between distinguishable particles, and one would need to resort to advanced numerical methods to solve recursive integral equations; with our effective Hamiltonian this step is trivial. We are also able to rearrange Eq. (\ref{delta0reduced}) for $\Delta_0$. Having obtained $\Delta_0$ and $\sumover{p}\Delta_\textbf{pq}$ in terms of $\chi_\textbf{q}$, we can use Eq. (\ref{chiDefinition}) obtain a single equation in terms of $\chi_\textbf{q}$
\begin{multline}
    \left(1-\frac{g_{AD}}{V}\sumover{p}\frac{{1}}{ E-\nu_\textbf{pq}-\frac{g_2^2}{V}\sumover{k}\frac{1}{E-\epsilon_\textbf{kpq}}}\right)\chi_\textbf{q}=\\
    \frac{g_{AD}}{V}\left( \frac{1}{E-\nu_0 -\frac{g^2_2}{V}\sumover{k}\frac{1}{E-2\epsilon_\textbf{k}}}\right)\left(\sumover{q}\chi_\textbf{q}\right).
\end{multline}
By observing that the rhs is constant with respect to the index $\textbf{q}$, we can rearrange one last time for $\chi_\textbf{q}$ and then take the sum over $\textbf{q}$ to obtain a single equation in terms of $\sumover{q}\chi_\textbf{q}$. Non-trivial solutions to this equation only exist if $E$ satisfies
\begin{multline}
    E-\nu_0 -\frac{g^2_2}{V}\sumover{k}\frac{1}{E-2\epsilon_\textbf{k}}= \\
    \frac{1}{V}\sumover{q}\frac{1}{\frac{1}{g_{AD}}-\frac{1}{V}\sumover{p}\frac{1}{E-\nu_\textbf{pq}-\frac{g^2_2}{V}\sumover{k}\frac{1}{E-\epsilon_\textbf{kpq}}}}.
    \label{DimerMedEqn}
\end{multline}
By running the coupling constants $g_2$, $\nu_0$ and $g_{AD}$ so that they reproduce vacuum two and three-body observables, this equation can be compactly written as
\begin{equation} \label{compactT2medeqn}
   \frac{8\pi}{R}\big[T_2(E)\big]^{-1}=\frac{\mu}{2\pi}\frac{1}{V}\sumover{q}\tilde{T}_{AD}\left(E+\epsilon_\textbf{q},\textbf{q}\right),
\end{equation}
where $T_2(E)$ is the two-body $T$-matrix (at rest) in vacuum and $\tilde{T}_{AD}(E+\epsilon_\textbf{q},\textbf{q})$ is the atom-dimer $T$-matrix in medium at finite momentum $\textbf{q}$, with an energy shift of $\epsilon_\textbf{q}$ to account for the energy required to create a hole of momentum $\textbf{q}$ in the Fermi sea. Eq. (\ref{compactT2medeqn}) is equivalent to finding the pole of the two-body scattering matrix in medium $\tilde{T}_2(E)$ obtained summing the Feynman diagrams presented in Fig. \ref{fig:T2Med diagrams}. Since we have ensured that all $T$-matrices are finite, and since $\sumover{q}$ is restricted to $|\textbf{q}|<k_F$, we can be sure that all quantities present in Eq. (\ref{compactT2medeqn}) are finite and require no further renormalization. We also observe that by taking $k_F\to0$ we recover the binding energy equation for a dimer in vacuum. In vacuum, Galilean invariance ensures that in vacuum the atom-dimer $T$-matrix at finite momentum $\textbf{q}$ is given by a simple energy shift ($T_{AD}(E,\textbf{q})=T_{AD}(E-\frac{1}{3}\epsilon_\textbf{q},0)$), however in medium the presence of the Fermi sea introduces a special reference frame making this approach invalid. One can obtain the analytic expression for $\tilde{T}_{AD}(E+\epsilon_\textbf{q},\textbf{q})$ by directly converting the sum into an integral, however the integrand is no longer spherically symmetric in $\textbf{p}$-space due to the term $\textbf{p}.\textbf{q}=|\textbf{p}||\textbf{q}|\cos(\theta_{pq})$. We eliminate this term by performing a coordinate transformation $\textbf{p}=\textbf{p}'+\textbf{q}/3$ yielding
\begin{multline}
    \big[\tilde{T}_{AD}(E+\epsilon_\textbf{q},\textbf{q})\big]^{-1}=\frac{1}{g_{AD}}\\
    -\frac{R}{8\pi}\frac{1}{V}\sumover{p'}T_2\left(E-\frac{1}{2\mu}(\textbf{p}')^2+\frac{2}{3}\textbf{q}^2\right), \label{TADMedFiniteQ}
\end{multline}
where $|\textbf{p}'+\textbf{q}/3|>k_F$ due to Pauli blocking and $T_2$ is the two-atom $T$-matrix in vacuum. With this coordinate substitution we trade a spherically symmetric integrand for more complicated integration bounds which encode the Pauli blocking. Both give the same result, however Eq. (\ref{TADMedFiniteQ}) is more convenient for analysing the behaviour of $\tilde{T}_2$ in the asymptotic limits $1/a_{AD}\to\pm\infty$. The results for the two-body binding energy in medium, corresponding to the dotted lines in Fig. \ref{fig:DimerTrimerMedium}, are obtained by first converting Eq. (\ref{TADMedFiniteQ}) to cylindrical coordinates which yields an analytic expression, before finally evaluating $\sumover{q}$ which must be performed numerically. The results in Fig. \ref{fig:DimerTrimerMedium} include a final energy shift by $+k_F^2/2$ to allow for direct comparison with the trimer results. We find that, similarly to the trimer, we are able to obtain solutions for any value of $1/a_{AD}$. Notably, irrespective of the value of $a_{AD}$ chosen, we find that the resulting binding energy is always below the vacuum binding energy ($E=-\kappa_+^2+k_F^2/2$).

\subsubsection{Case $\frac{1}{a_{AD}}\to-\infty$}
We begin by defining $|\textbf{q}|=q$ . For $q<k_F$ the Pauli blocked region in Eq. (\ref{TADMedFiniteQ}) is a sphere in $\textbf{p}'$-space centred at $\textbf{p}'=-\textbf{q}/3$ and radius $k_F$. Since the integrand is spherically symmetric, we are free to align our $z$-axis in $\textbf{p}'$-space to be along $\textbf{q}$. For all $q<k_F$, the smallest value of $|\textbf{p}'| $ which is not in the excluded region is found at the top of the sphere at $\textbf{p}'=(0,0,k_F-q/3)$. In order to not integrate over a pole in $\textbf{p}'$-space, we need to ensure that $(\textbf{p}')^2/(2\mu)-E-1/3q^2>\kappa_+^2$ . Together, this means that for finite $q$ the highest energy that guarantees we do not integrate over a pole is
\begin{equation}
    E<-\kappa_+^2+\frac{1}{2\mu}\bigg(k_F-q\bigg)\bigg(k_F+\frac{1}{3}q\bigg).
\end{equation}
Since we are integrating over the interval $0<q<k_F$, we must take $E$ to be lower than the lowest bound, which is $E<-\kappa_+^2$ and occurs as $q=k_F$. Provided that the pole of $\tilde{T}_{AD}(E+\epsilon_{k_F},k_F)$ occurs above $E=-\kappa_+^2$ for $1/a_{AD}\to-\infty$, this guarantees that $\sumover{q}\tilde{T}_{AD}(E+\epsilon_\textbf{q},\textbf{q})\to0^-$ as it involves integrating over a small but finite quantity. Consequently, the lhs of Eq. (\ref{compactT2medeqn}) must also approach this limit, which necessitates $E\to-\kappa_+^2$. This analysis suggests that the role of Pauli blocking becomes negligible in this asymptotic limit and the dressing of the dimer by p-h excitations is suppressed, recovering the vacuum dimer binding energy. Our results support this view, with all dimer solutions tending to $E=-\kappa_-^2+k_F^2/2$ once the energy shift which allows for comparison with trimer solutions is included.

\subsubsection{Case $\frac{1}{a_{AD}}\to+\infty$}
Once again, we need to ensure that we are not integrating over a pole for any value of $q$. Previously we were focused on not encountering a pole when taking $\sumover{p'}$ which necessitated $E<-\kappa_+^2$, however for positive values of $1/a_{AD}$ our primary concern is not encountering the pole of $\tilde{T}_{AD}(E+\epsilon_\textbf{q},\textbf{q})$. Following our analysis of the trimer problem in medium, we know that for $q=0$ there will be a pole and that this pole approaches the vacuum trimer pole for large values of $1/a_{AD}$, which is significantly lower than the vacuum dimer pole at $E=-\kappa_+^2$. Since the contribution of Pauli blocking is negligible for such energies, we expect that at finite momenta the poles of  $\tilde{T}_{AD}(E+\epsilon_\textbf{q},\textbf{q})$ approach the vacuum results, therefore we can approximate
\begin{equation}
    \tilde{T}_{AD}(E+\epsilon_\textbf{q},\textbf{q})\approx T_{AD}\left(E+\frac{2}{3}\epsilon_\textbf{q}\right)
\end{equation}
If we denote the vacuum binding energy for a trimer as $E_T$, then the upper bound of $E$ that ensures we do encounter a pole for any $q$ in $\sumover{q}$ is $E<E_T-2\epsilon_{k_F}/3$. For such low energies, significantly below the vacuum dimer binding energy, we must have that the lhs of Eq. (\ref{compactT2medeqn}) is negative as the dominant contribution is linear in $E$. Since $\sumover{q}$ is finite, the only way for this sum to produce such a significant contribution and remain negative is if the energy $E$ approaches $ E_T-2\epsilon_{k_F}/3$ from below. 

\begin{figure}[t]
    \centering
    \includegraphics[width=\linewidth]{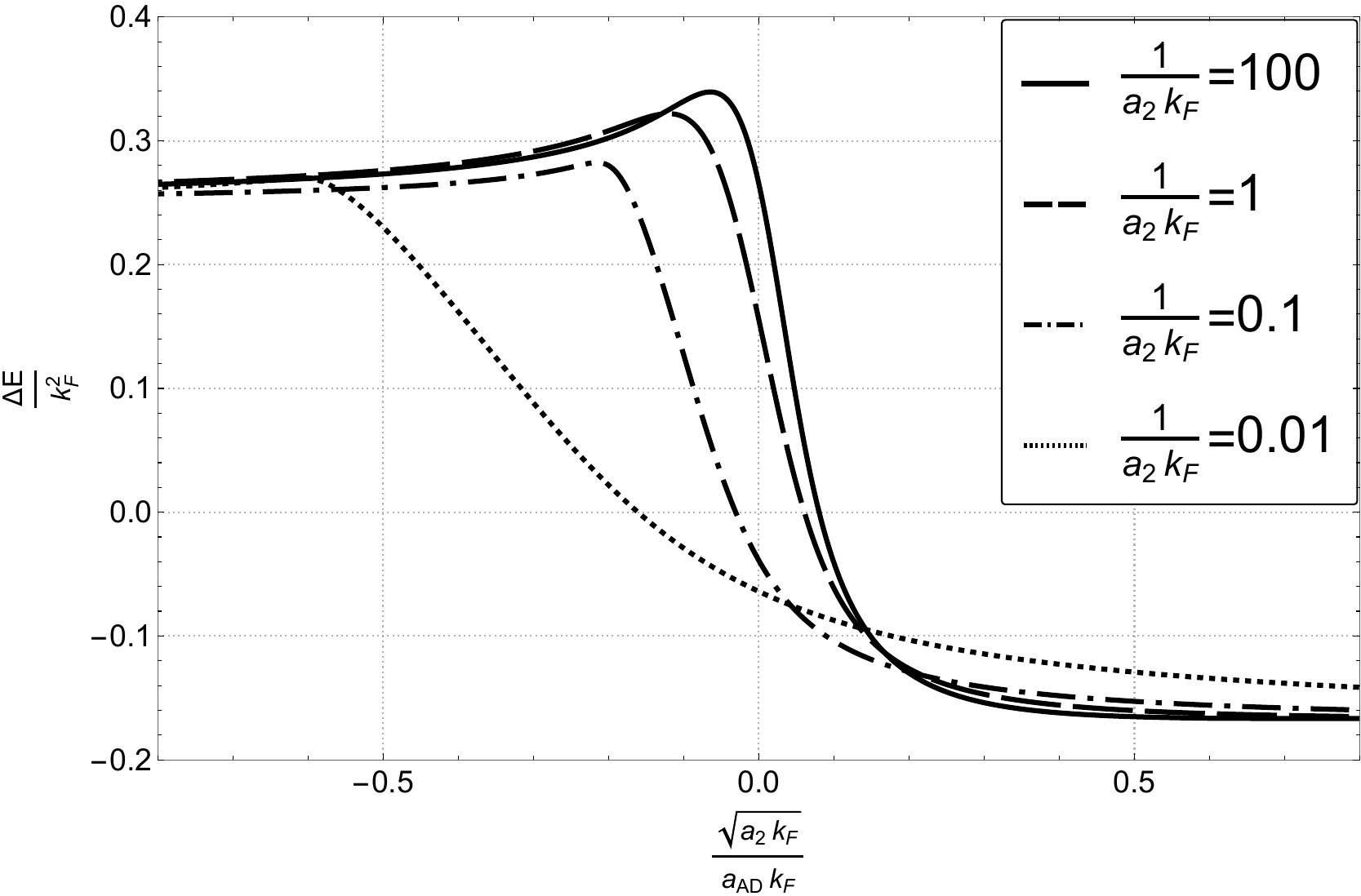}
    \caption{Plot of the difference in energy between trimer and dimer branches ($\Delta E/k_F^2$) for several values of $1/a_2k_F$ spanning several orders of magnitude.}
    \label{fig:TrimerDimerDiff}
\end{figure}

\subsection{Crossover}
The results presented in Fig. \ref{fig:DimerTrimerMedium} highlight the key features of the dimer and trimer solutions as $1/a_2$ is varied across several orders of magnitude. Notably, all results demonstrate a crossover from a dimer to a trimer with increasing values of $1/a_{AD}$. To highlight this crossover, we also present Fig. \ref{fig:TrimerDimerDiff} which shows the energy difference between the trimer and dimer branches ($\Delta E/k_F^2$) as $1/a_{AD}$ is varied while keeping all two-body observables constant. We remark that the scaling of the horizontal axis by a factor $\sqrt{a_2k_F}$  in both of these figures is not particularly significant and has been chosen primarily to elucidate the crossover as $1/a_2$ changes over a wide range; it is likely related to the prefactor $(\kappa_+-\kappa_-)^{-1}=R/\sqrt{1+R/a_2}$ (as seen in the rhs of Eq. (\ref{TADVacuumAnalyticExpression}) and Eq. (\ref{TADMedAnalytic}). The fact that a crossover is guaranteed for any value of $1/a_2$ can be deduced from our earlier analysis of the trimer and dimer binding equations in medium in the limits $1/a_{AD}\to\pm\infty$. Once particle number has been accounted for, we can expect the trimer (dimer) binding energy to tend to $-\kappa_+^2+3k_F^2/4$ ($-\kappa_+^2+k_F^2/2$) when $1/a_{AD}\to-\infty$ and $E_T$ ($E_T-k_F^2/6$) when $1/a_{AD}\to+\infty$, where $E_T$ denotes the trimer energy in vacuum. The energy difference between the trimer and dimer solutions is therefore expected to be $\Delta E/k_F^2\to1/4$ (implying the dimer is the ground state) and $\Delta E/k_F^2\to-1/6$ (trimer is the ground state) for $1/a_{AD}\to-\infty$ and $1/a_{AD}\to+\infty$ respectively, which is confirmed by the results presented in Fig. \ref{fig:TrimerDimerDiff}. While the existence of a crossover is therefore guaranteed, its exact position is not fixed. The results of Fig. \ref{fig:TrimerDimerDiff} show that as the binding within the dimer decreases ($1/a_2\to0^+$) the trimer branch becomes energetically favourable for $1/a_{AD}$ closer to unitarity and eventually even becomes favoured over the dimer in a region where $1/a_{AD}<0$. In the case of the familiar polaron-molecule transition, one expects to find a crossover only for positive values of the scattering length, meaning that the ground state can either be a dressed impurity or a localised molecule. Our results suggest that for sufficiently small dimer binding energy one is able to crossover from a dressed dimer state directly to a Cooper pair formed of a dimer and a $c$-particle, with the localised trimer only becoming the ground state when $1/a_{AD}>0.$

\section{Summary and conclusions}
In conclusion, we have studied the interaction between a (non-universal) dimer formed of two distinguishable particles and a third distinguishable (Fermionic) species. To this end, we have introduced an effective Hamiltonian and demonstrated that is is renormalizable in the case where three particles are in vacuum. In the process, we have seen how such a system yields remarkably simple binding equations and avoids the pathologies that are often encountered when studying the three-body problem. We are able to solve these equations, and we showed that in the limit where the dimer becomes deeply bound the three-body problem reduces to a two-body problem between particles of unequal masses interacting through a contact interaction. We have also investigated the opposite limit where the dimer binding energy approached the two-atom threshold and we have confirmed that the theory remains renormalizable in this limit. We then considered placing the dimer in a Fermi sea formed by the third species and studied the competition between a dimer dressed by p-h excitations with the medium and a trimer subject to Pauli blocking. Much like with the polaron-molecule transition, we found that by varying the atom-dimer scattering length the system is able to transition between states where the dressed dimer is the ground state and where the trimer is the ground state. By varying the binding energy of the dimer, we are also able to control the position of this crossover. This additional degree of freedom is not present when studying the usual polaron-molecule transition, and our results show that by decreasing the dimer binding energy the crossover can occur for negative values of $1/a_{AD}$, opening up the possibility of observing a Cooper pair between a dimer and a third atom as the ground state of the system.

\nocite{*}

\section{Acknowledgements}

FC acknowledges support from Institut Universitaire de France, PEPR Project Dyn-1D (ANR-23-PETQ-001), ANR Collaborative project ANR-25-CE47-6400 LowDCertif and ANR International Project ANR-24-CE97-0007 QSOFT. 

\bibliography{apssamp}

\end{document}